\documentclass[reprint,aps,amsmath,amsfonts,sort&compress,superscriptaddress,noeprint,amssymb,floatfix]{revtex4-2}  

\usepackage[breaklinks=true,colorlinks=true,citecolor=blue,urlcolor=blue,linkcolor=blue,pdfencoding=auto,psdextra]{hyperref}
\usepackage{balance}

\usepackage{amsmath,amssymb,amstext}
\usepackage{subfigure}

\usepackage{float}
\usepackage[version=4]{mhchem}
\usepackage{multirow}
\usepackage[utf8]{inputenc}
\usepackage{lmodern}

\usepackage{mathtools,nccmath}%

\usepackage{environ}

\usepackage{xcolor}

\newcommand{\s}{\,s$^{-1}$}
\newcommand{\ergs}{\mbox{\,erg\,\s}}

\newcommand{\rc}{r_{\rm c}}
\newcommand{\lx}{L_{\rm x}}

\newcommand{\rl}{\mathcal{R}_L}
\newcommand{\lc}{\ell_{\rm c}}
\newcommand{\mbh}{M_\bullet}

\newcommand{\lbol}{L_{\rm bol}}

\def\app#1#2{%
  \mathrel{%
    \setbox0=\hbox{$#1\sim$}%
    \setbox2=\hbox{%
      \rlap{\hbox{$#1\propto$}}%
      \lower1.1\ht0\box0%
    }%
    \raise0.25\ht2\box2%
  }%
}

\begin{document}

\title{\large Particle Acceleration, Coronal Neutrino Production, and the Diffuse Extragalactic Neutrino Background from Supermassive Black Holes}
\author{Rostom Mbarek}
\email{rmbarek@princeton.edu}
\affiliation{Department of Astrophysical Sciences, Princeton University, Princeton, NJ 08544, USA}

\begin{abstract}
We present a generalized neutrino luminosity function for protons accelerated in the X-ray coronae of supermassive black holes in Seyfert-like galaxies. A major uncertainty in assessing the diffuse neutrino contribution of these systems is the underlying particle acceleration physics. We address this using a theoretical acceleration framework informed by plasma kinetic simulations, enabling a more self-consistent connection between coronal conditions, nonthermal proton populations, and neutrino production. In this picture, the neutrino luminosity depends primarily on the coronal X-ray luminosity and magnetization, and only weakly on black hole mass. We find that the cosmologically integrated emission from these systems can account for the sub-PeV diffuse extragalactic neutrino flux observed by IceCube. We further argue that, although diffusive confinement is relatively well understood, the magnetic field topology near black holes naturally allows for cosmic ray-driven outflows near the X-ray corona. Such outflows may accompany additional efficient neutrino production at the PeV-level and influence the dynamics of the innermost galactic environment.

\end{abstract}

\maketitle
\defcitealias{mbarek+24}{M24}

\section{Introduction}\label{sec:intro}

A diffuse neutrino excess beyond the atmospheric background has been steadily established by IceCube over the past decade in the TeV--PeV range \citep[e.g.,][]{ICECUBE15b,ICECUBE16}. Its approximate isotropy points to an extragalactic origin, and its energy flux is comparable to that of $\gamma$-rays in the local universe \citep{ackermann+15,fang+18}. More recently, IceCube has reported a significant neutrino flux from the direction of the Seyfert galaxy NGC~1068 \citep{IceCube-NGC1068}. The lack of a commensurate GeV--TeV $\gamma$-ray counterpart \citep{aartsen+20,IceCube-NGC1068} suggests that the source is electromagnetically hidden, with the accompanying cascade power likely reprocessed to the MeV band \citep{murase+20b,ajello+23}. These observations motivate highly magnetized X-ray coronae around accreting supermassive black holes as promising sites of TeV--sub-PeV neutrino production \citep[e.g.,][]{murase+20b,inoue+20,kheirandish+21,murase22,halzen+22,kurahashi+22,eichmann+22,mbarek+24,fiorillo+24,padovani+24,testagrossa+26}, and as possible contributors to the diffuse sub-PeV neutrino background \citep[e.g.,][]{padovani+24b,fiorillo+25,karavola+26,
saurenhaus+26,murase+26,carpio+26,yang+26}.

The relevant physical picture builds on two ingredients. First, nonthermal protons can be stochastically accelerated in turbulent, magnetized coronae with spectral tails regulated by the turbulence and magnetization \citep{murase+20b,mbarek+24,fiorillo+24b,lemoine+25,lebihan+26,mbarek+26a}. Second, these protons interact primarily with coronal X-ray and disk UV photons, producing neutrinos and $\gamma$-rays through photomeson processes \citep{mbarek+24}. The resulting neutrino luminosity is controlled mainly by the X-ray luminosity $L_{\rm x}$, which sets the thermal coronal conditions \citep{mbarek+24} and traces the associated UV radiation field \citep[e.g.,][]{lusso+10}, and more weakly by the SMBH mass $M_\bullet$, which sets the characteristic coronal size \citep[e.g.,][]{fabian12,miller+13,fabian+09,dai+10,demarco+11,kara+13} and hence the target-photon energy density.

In this paper, we show that proton acceleration in the X-ray coronae of Seyfert-like galaxies, modeled using the magnetized-turbulence scaling of \citet{mbarek+26a}, can reproduce the sub-PeV diffuse neutrino flux observed by IceCube. We derive source-specific coronal proton distributions and compute the corresponding neutrino output, showing that the emission remains consistent with current observational limits while allowing X-ray coronae to account for a substantial fraction, and potentially all, of the measured sub-PeV flux. This supports a picture in which Seyfert galaxies and quasars form a population of steady sub-PeV neutrino emitters, with the brightest nearby systems potentially resolvable in current or next-generation neutrino data.

Several sources already provide promising tests of this picture. NGC~4151 and NGC~3079 may be detectable with decade-scale IceCube exposure \citep{neronov+24}; NGC~4151 and CGCG~420-015 show a $2.7\sigma$ excess in an IceCube search of X-ray-bright northern Seyferts \citep{ICECUBE_25b}; NGC~4945 and Circinus are among the most promising southern targets for KM3NeT, Baikal-GVD, and related detectors in disk-corona models \citep{murase+24}; and NGC~7469 has been discussed in connection with two $\sim100\,{\rm TeV}$ IceCube alerts \citep[e.g.,][]{sommani+25,eichmann+26}. Recent IceCube searches also report collective evidence for X-ray-bright Seyfert emission in both northern and southern samples \citep{ICECUBE_25b,ICECUBE26a}.

The large proton luminosities required by the neutrino signal also imply that a non-negligible fraction of the accelerated protons could potentially escape the corona along the expected open magnetic-field topology near the black hole. 
This escaping component could power a cosmic-ray-driven outflow whose energetics can be estimated quantitatively. We conclude that the coronal cosmic-ray power, while likely not exceeding the X-ray luminosity \citep{murase22,das+24}, is sufficient to account for the observed sub-PeV diffuse neutrino emission.

\section{A Source-Dependent Neutrino Luminosity} \label{sec:GNL}

We express the proton energy density necessary to produce neutrinos as $E_p^2 \frac{dn_p}{dE_p} \approx \frac{L_p}{4 \pi \rc^2 c}$, where $L_p$ is the proton luminosity in the coronal source, and $\rc$ is the coronal size, which is consistently found to be $\rc \simeq 10\, r_g = 10\, GM_\bullet/c^2$, where $M_\bullet$ is the black hole mass \citep[e.g.,][]{dai+10,fabian12,fabian+15,wilkins+21}. 
The ensuing neutrino luminosity $L_\nu$ from the dominant neutrino production route, i.e., photomeson interactions ($p\gamma$) is \citep{mbarek+24}
\begin{equation}\label{eq:Lnu1}
	L_\nu(\alpha E_p) \simeq  \frac{3 \pi }{2} c r_c^2 \kappa_{p \gamma}   E_p^2 \frac{dn_p}{dE_p}
\end{equation}
such that $E_p$ is the proton energy, $\alpha \simeq 1/20 = E_\nu/E_p$ where $E_\nu$ is the neutrino energy, $\kappa_{p\gamma}(E_p) = \text{min}(t^{-1}_{p\gamma} t_{\rm esc} , 1)$ is the neutrino production efficiency where $t_{p\gamma}$ is the proton cooling time, and $t_{\rm esc}$ is the proton escape time from the corona.
The rate $t^{-1}_{p\gamma}$ is expressed as  $t^{-1}_{p\gamma}  = \xi \sigma_{\rm eff} \frac{L}{6 \pi \rc^2 \epsilon_0}$, where $\xi = 0.17$ is the inelasticity, $\sigma_{\rm eff} \simeq 250 \mu$b is an effective cross section for $p\gamma$ interactions \footnote{This is a fit to the cooling time calculated based on an energy-dependent cross section \citep{PDG18}}, $L$ the luminosity of the radiation source, and $\epsilon_0$ the dominant photon energy.
In the following, we generalize Eq.~\eqref{eq:Lnu1}, by gathering a source-dependent \emph{i)} proton energy density $E_p^2 \frac{dn_p}{dE_p}$, \emph{ii)} escape time $t_{\rm esc}$, and \emph{iii)} $p\gamma$ cooling time $t^{-1}_{p\gamma}$.

\subsection{ Generalized Proton Energy Density}\label{sec:p-density}

The density of nonthermal protons that produce neutrinos in X-ray coronae scales with the bulk density of coronal protons $\bar{n}_p$ \citep{mbarek+24}. The tail is normalized at the injection scale $\gamma_p\simeq\sigma_p$, so $\gamma_p/\sigma_p$ measures the corresponding Larmor-radius separation from injection. We can express the spectral slope of stochastically-accelerated particles in regions with $\delta B/B \sim 1$ as $n_p(\gamma_p) \simeq \gamma_p \frac{dn_p}{d\gamma_p} = \bar{n}_p \left( \frac{\gamma_p}{\sigma_p} \right)^{-s+1}$ \citep{mbarek+26a}:
\begin{equation}\label{eq:slope}
   s \simeq 2 + \frac{\pi}{2} \sqrt{\frac{1 + \sigma_p}{\sigma_p}} \left(\frac{\gamma_p \frac{m_p}{m_e} d_e}{\sqrt{\sigma_e} \lc} \right)^{r} \frac{1}{\ln \mathcal{G}}
\end{equation}
Here, $\sigma_e = \frac{B^2}{4 \pi \bar{n}_e m_e c^2}$ is the pair magnetization and $\sigma_p = \frac{B^2}{4 \pi \bar{n}_p m_p c^2}\geq 0.1$ \citep{mbarek+26a} is the proton magnetization, such that $\bar{n}_p(m_p)$ and $\bar{n}_e(m_e)$ are the proton, and electron bulk density(mass), respectively. This expression is relevant for ion-electron plasma and pair plasma with a subdominant population of ions.
The term $d_e$ is the skin depth of the plasma, $\mathcal{G} =1+ \Delta \gamma/\gamma$ is the energy gain per cycle such that $\mathcal{G} \simeq 1 + 4/\pi \sqrt{\sigma_p/(1+\sigma_p)}$ for mirror acceleration \citep{vega+24,mbarek+26a}, $\lc = \zeta r_g$ is the coherence length of the magnetic field with $\zeta \sim 1$ \footnote{The value of $\zeta$ is likely source-dependent, but expected to remain of order unity \citep{mbarek+24,mbarek+26a}. Pinning down its precise value will require robust localization of the corona in global GRMHD simulations.}.

The exponent $r$ describes the scale dependence of the scattering and acceleration times. For Fermi acceleration, $(\Delta\gamma)^2\propto\gamma^2$, while the interval between interactions is the scattering time, $t_{\rm sc}\propto \rl^r$, where $\rl$ is the proton Larmor radius. Consequently, the diffusion coefficient $D_{\gamma\gamma} = \gamma^2/t_{\rm acc} \propto\gamma^{2-r}$ and $t_{\rm acc}=\gamma^2/D_{\gamma\gamma}\propto\gamma^r$. In this paper, we adopt $r\simeq0.3$ as $\rl \to \lc$ \citep{lemoine23,kempski+23,mbarek+26a}, as inferred from the field-line-curvature statistics in the driven-turbulence kinetic simulations of \citet{mbarek+26a}. An energy-independent acceleration time corresponds to $r=0$. At the level of the energy scaling, this is equivalent to the $q=2$ hard-sphere limit for which $t_{\rm acc}\propto E_p^{2-q}$ \citep[e.g.,][]{murase+20b,murase+26}.

We can express the coronal electron density as $\bar{n}_e \simeq \tau/(\sigma_T r_c)$, where $\tau \simeq 1$ \citep[e.g.,][]{rybicki+79,fabian+15,beloborodov17} is the optical depth and $\sigma_T$ is the Thomson cross-section. 
The pair magnetization is further expressed as $\sigma_e = \frac{2 \ell }{\tau } \frac{U_{\rm B}}{U_{\rm x}}$, where $\ell=\sigma_T U_{\rm x} r_c/m_e c^2$ is the radiative compactness, and $U_{\rm x} = \lx/4 \pi c \rc^2$ and $U_{\rm B} = B^2/8 \pi$ are the X-ray and magnetic energy density, respectively. In a turbulent corona scenario, $U_{\rm x} \sim U_{\rm B}$ \citep[see Appendix~\ref{app:uxub} and][]{groselj+24}, so we can rewrite $\sigma_{e}\approx  2\ell/\tau$.
Now we can express the proton bulk density as $\bar{n}_p  \sigma_p = \sigma_{e} \frac{m_e \bar{n}_e}{m_p}$, so we can rewrite $\bar{n}_p \simeq 2U_{\rm x}/(\sigma_p m_p c^2)$, and the energy density of protons in erg~cm$^{-3}$ is,
\begin{equation}\label{eq:pden1}
\begin{split}
    E_p^2 \frac{dn_p}{dE_p} (\gamma_p \geq 10^4 , \sigma_p, \lx, \mbh) \simeq \frac{\lx}{ 2 \pi \sigma_p c \rc^2} \left( \frac{\gamma_p}{\sigma_p} \right)^{-s + 2}
    \\
    \simeq 2.5 \times 10^{6} \frac{\lx}{10^{44} \sigma_p \rm \ergs} \left( \frac{10^7 M_\odot}{M_\bullet} \right)^{2} \left( \frac{\gamma_p}{\sigma_p} \right)^{-s + 2}
\end{split}
\end{equation}
where the proton distribution is constructed for $\gamma_p \geq 10^4 \gg \sigma_p $, which ensures that $\rl$ approaches the turbulence coherence length ($\lc \sim r_g \propto \mbh$ \citep{mbarek+24}) even for large black hole masses. This normalization places the nonthermal proton energy density at $U_p\sim U_{\rm B}$ (Appendix~A in \citep{groselj+24}), i.e., $L_p\lesssim \lx$, so the power carried by the neutrino-producing protons lies within an order-unity-fraction limit of the turbulent magnetic dissipation that also powers the corona. In this regime, turbulent structures with $\delta B/B \sim 1$ which are required for efficient stochastic acceleration, are expected to develop \citep{mbarek+26a}. More generally, the closer the proton population extends toward the maximum achievable Lorentz factor, $\gamma_{\rm max}$, for which $\rl \sim \lc$, the more robust the resulting power-law predictions become. 

To build intuition for the proton distribution in Eq.~\eqref{eq:pden1}, we revisit Eq.~\eqref{eq:slope} and simplify it by adopting the likely value of $\sigma_p \sim 1$ for turbulent acceleration \citep{mbarek+24}. We then rewrite the electron skin depth as $d_e = \sqrt{ \frac{m_e c^2}{4 \pi \bar{n}_e e^2}} = \sqrt{ \frac{10 \sigma_T m_e c^2}{4 \pi e^2 \tau}} \sqrt{r_g}$, and then express the magnetization $\sigma_{e}\approx  2\ell/\tau = \frac{\sigma_T \lx }{20 \pi r_g m_ec^3 \tau}$. 
With these substitutions, the spectral slope becomes,
\begin{equation}\label{eq:slope2}
    s \simeq 2 + 3.5 \left(3 \times 10^{-10}  \frac{\gamma_p}{\zeta}  \sqrt{\frac{\rm 10^{44} erg/s}{\lx}} \right)^r
\end{equation}
where $(s-2) \propto (\gamma_p/\zeta)^r\lx^{-r/2}$, depending mainly on $\lx$. We emphasize that Eq.~\eqref{eq:slope2} is an approximation of Eq.~\eqref{eq:slope}.

The highest-energy protons should arise primarily from high-luminosity AGN, especially quasars with $\lx \gtrsim 10^{45} \ergs$ \citep{saccheo+23}. This is because (i) the proton normalization in Eq.~\eqref{eq:pden1} scales with the coronal X-ray energy density $\propto \lx/\mbh^2$, and (ii) more luminous sources can sustain harder proton spectra up to higher energies (see Eq.~\eqref{eq:slope2})--because in this fast-acceleration scenario ($t_{\rm acc} \ll t_{\rm cool}$; see Bethe--Heitler
Suppression) the maximum energy is set by confinement ($\rl \to\lc$) rather than by cooling, so the
harder slope of luminous sources is not offset by an earlier cooling cutoff. However, this normalization is non-trivial: it decreases with increasing $M_\bullet$, which often correlates with higher $\lx$ (see Appendix~\ref{app:rholx}), and also depends on $\sigma_p$. In NGC~1068, we find $\sigma_p \sim 1$ \citep{mbarek+24}, although this value may vary across sources. Future neutrino detections, together with kinetic simulations of nonthermal spectra in turbulent media, will be important for constraining $\sigma_p$.

\subsection{Generalized Cooling time}
The generalized cooling time from $p \gamma$ interactions between X-rays and protons is expressed as $t^{-1}_{p\gamma}  = \xi \sigma_{\rm eff} \frac{L_{\rm x}}{6 \pi \rc^2 \epsilon_0}$ \citep{mbarek+24}, such that $\epsilon_0 \simeq 7$keV is defined in terms of the iron K$\alpha$ line \citep[e.g.,][]{reynolds+99} where a large fraction of the X-ray power lies. Here $L_{\mathrm{x}}$ denotes the intrinsic 2--10~keV coronal X-ray luminosity. Because the present calculation integrates over a population of X-ray sources, each with its own intrinsic spectral shape, assigning a detailed X-ray slope on a source-by-source basis is not well constrained. We approximate the X-ray target at $\epsilon_0 \simeq 7$~keV, near the Fe~K$\alpha$ reprocessing feature where much of the reprocessed power density resides \citep{mbarek+24}. The intrinsic coronal X-ray field is a broad Comptonized continuum. The characteristic energy $\epsilon_0\simeq7$~keV is a population-level coarse graining, not a monochromatic-corona assumption. A more realistic spectrum can modify the photohadronic interaction rate, particularly for individual non-calorimetric sources and near the interaction threshold. This spectral dependence is therefore an important limitation in those regimes. However, it should have less of an impact on the integrated normalization from the luminous AGN that dominate the diffuse result below.

Beyond X-rays, higher energy protons additionally interact with lower energy radiation (e.g., Ultra-violet (UV) light) to produce neutrinos, all within the corona.
We can then rewrite $t^{-1}_{p\gamma}$ as \citep{mbarek+24},
\begin{equation}\label{eq:tpg}
	t^{-1}_{p \gamma}(\gamma_p) \frac{6 \pi \rc^2}{ \xi \sigma_{\rm eff}}  = 
	\begin{cases}
		\frac{\lx}{\epsilon_0},& \text{if }  \frac{\bar{\epsilon}_{\rm th}}{2 \epsilon_0} < \gamma_p < \frac{\bar{\epsilon}_{\rm th}}{2 \epsilon_1} \\
		\frac{\lx}{\epsilon_0} + \frac{L_{\rm UV}}{(r_{\rm UV}/r_c)^2\,\epsilon_1} & \text{if } \frac{\bar{\epsilon}_{\rm th}}{2 \epsilon_1} < \gamma_p 
	\end{cases}	
\end{equation}
where $\epsilon_1 < \epsilon_0$, $r_{\rm UV}$ is the half-light radius of the corresponding radiation component, and $\bar{\epsilon}_{\rm th} = 0.15$~GeV is the threshold for $p\gamma$ interactions in the proton frame.

Different radiation components that peak at characteristic energies $\epsilon_i$ can be included depending on the dominant radiation fields. We focus on UV emission because i) it is the primary radiation component in the inner regions of AGN disks \citep[e.g.,][]{edelson+15,de_rosa+15,cackett+21}, and
ii) photomeson interactions between relativistic protons and UV photons produce neutrinos with characteristic energies in the diffuse energy band
$E_\nu \simeq \alpha \left( \frac{\bar{\epsilon}_{\rm th}}{2 \epsilon_{\rm UV}} \right) m_p c^2 \gtrsim 10^{14}$eV, where $\epsilon_{\rm UV} \sim 10$eV is the energy band where UV power is concentrated \citep[e.g.,][]{shull+12}.

\subsection{Bethe-Heitler Suppression}\label{sec:BHe}

The density of protons available for $ p\gamma $ interactions is suppressed due to the Bethe-Heitler (BHe) process ($ p + \gamma \rightarrow p e^+ e^- $), where protons interact with photons to produce pairs in coronae \citep{murase+20b}. To quantify this effect, we define a suppression factor based on the number of protons undergoing $ p\gamma $ interactions, $ N_{p\gamma} $, which scales as $ N_{p\gamma} \propto t_{p\gamma}^{-1} $. Similarly, the number of protons undergoing BHe interactions follows $ N_{\rm BHe} \propto t_{\rm BHe}^{-1} $. The effective density of protons available for $ p\gamma $ interactions is then adjusted by a BHe suppression factor,  $
f_{\rm BHe} = \frac{N_{p\gamma}}{N_{p\gamma} + N_{\rm BHe}} = \frac{t_{\rm BHe}}{t_{p\gamma} + t_{\rm BHe}}
$ (see Appendix~\ref{app:bh}).
We note that BHe losses play a different role in our calculation than in \citet{murase+20b}, owing to the different acceleration time adopted here \citep{mbarek+26a}. In \citet{murase+20b}, stochastic acceleration is slow enough for BHe cooling on dense disk UV photons to compete with acceleration and limit the proton spectrum at $\sim0.1$--$1$~PeV. In our framework, as in other PIC studies, tangible nonthermal power laws capable of reaching the proton energies required for high-energy neutrinos start arising in moderately magnetized collisionless  plasma with $\sigma_p\geq 0.1$ \citep[e.g.,][]{zhdankin+17,comisso+18,comisso+22}. This magnetization exceeds the $\sigma_p\sim10^{-2}$ expected for an ion--electron corona at virial $T_p$ with $\beta\sim1$ \citep{murase+20}, thereby shortening the acceleration time, $t_{\rm acc}\propto\sigma_p^{-1}$. Additionally, the acceleration time is decreased further in our model with the fiducial $r\simeq0.3$ (energy-dependent acceleration time).

Several PIC studies find $D_{\gamma \gamma}\propto \gamma^2$, corresponding to an approximately energy-independent acceleration time, over their resolved
high-energy nonthermal ranges \citep{comisso+19,wong+20,wong+25,gorbunov+25}. However, departures occur at lower energies \citep{wong+20} and near the Hillas limit \citep{gorbunov+25,wong+25}. Moreover, the weak dependence adopted here, $t_{\rm acc}\propto E^{0.3}$, changes the acceleration time by only a factor of $\simeq2$ per energy decade. Given the limited dynamical range
of current simulations, these results do not cleanly distinguish $r=0$ from $r\simeq0.3$. We therefore treat the energy dependence as model dependent rather than regarding either value as conclusively established.

The \citet{mbarek+26a} framework provides a physical basis for the weak
energy dependence adopted here by connecting the scattering time to the
scale-dependent statistics of intermittent field-line curvature. The
resulting $r\simeq0.3$ scaling is tested against stationary particle
spectra from driven electron--ion PIC simulations with particle escape
and thermal reinjection. 
For the same parameters at $\rl \lesssim \ell_c$, adopting $r=0$ makes
$t_{\rm acc}$ longer than our $r\simeq0.3$ prescription. The ordering
$t_{\rm acc}\ll t_{\rm BHe},t_{p\gamma}$ would not necessarily hold
throughout the acceleration range. For $r=0$, directly adopting the
\citet{mbarek+26a} steady-state proton spectrum would not be justified.

\subsection{Generalized Escape Time}\label{sec:escape}

Charged particles can in principle leave the corona by diffusion, streaming, or advection. However, large bulk losses are unlikely to dominate in the X-ray-emitting region. Losses on a light-crossing time $t_{\rm lc} = r_c/c$ would be difficult to reconcile with radiatively stable Seyfert coronae that have Thomson optical depths of order unity and produce standard X-ray spectra \citep[e.g.,][]{fabian+15,beloborodov17}. In addition, if non-diffusive escape removed most of the nonthermal proton population, the steady-state high-energy proton density would fall below that required to sustain the coronal neutrino signal inferred for NGC~1068 \citep{mbarek+24}. 

At the same time, a subdominant escaping component is well motivated. Hints of coronal outflows suggest that advection can carry some energetic particles outward \citep[e.g.,][]{hankla+26}, and the magnetic topology near accreting black holes, especially in MAD-like states relevant to the $\sigma_p\sim1$ regime considered here, can favor streaming along a dominantly poloidal guide field \citep[e.g.,][]{tchekhovskoy+10,liska+20}. Thus, non-diffusive transport can plausibly let a small fraction of ions escape on timescales shorter than the diffusive time without depleting the in-situ population. The coexistence of a long-lived corona with an observable neutrino population therefore constrains streaming or advective escape to be a leakage channel rather than the dominant sink. Characterizing the average proton population that produces the local neutrino signal then requires adopting diffusion in Eq.~\eqref{eq:Lnu1}.

We therefore take the effective escape time of the neutrino-producing population to be $t_{\rm esc}\simeq t_{\rm diff}$, with the diffusive escape time estimated as
$t_{\rm diff}\sim \rc^2/(\rl^r\lc^{1-r}c)$, where $\rl$ is the Larmor radius \citep{mbarek+24}. Using the coronal field implied by $U_{\rm x}\sim U_{\rm B}$ gives
$\rl\simeq (m_pc^2/e)(\gamma_p/\sqrt{8\pi U_{\rm x}})$, and therefore
\begin{equation}\label{eq:tesc}
\begin{split}
    t_{\rm diff}(\gamma_p)  \simeq 10^{2 - r}   \left( \frac{e}{m_p} \right)^r \zeta^{r - 1} \left( \frac{2\lx}{c^5} \right)^{r/2} \left( \frac{r_g}{c} \right)  \gamma_p^{-r}  \\
    \simeq \frac{3 \times 10^6}{\rm s} \left( \frac{L_{\rm x}}{10^{44}\rm \ergs} \right)^{0.15} \left( \frac{M_\bullet}{10^7 M_\odot} \right) \gamma_p^{-0.3},
\end{split}
\end{equation}
where the second line uses the canonical value $r\simeq0.3$ and $\zeta=\lc/r_g=1$.

For NGC~1068, $r_c=10r_g$ and $M_\bullet\simeq10^7M_\odot$ give $t_{\rm lc}=r_c/c\simeq5\times10^2\,{\rm s}$, whereas
Eq.~\eqref{eq:tesc} gives $t_{\rm diff}\simeq4\times10^4$--$2\times10^5\,{\rm s}$ for $\gamma_p\simeq10^4$--$10^6$. Thus prompt non-diffusive escape would reduce the residence time by $t_{\rm lc}/t_{\rm diff}\sim3\times10^{-3}$--$10^{-2}$, depending mostly on $\sim \lc/r_c$. The neutrino-producing protons must therefore be confined for many light-crossing times on average. The ratio $t_{\rm lc}/t_{\rm diff}$ sets how much faster than diffusion any escape channel can act before invalidating the diffusive estimate of the proton density and thereby Eqs.~\eqref{eq:Lnu1} and \eqref{eq:pden1}.

\subsection{Generalized Neutrino Luminosity}\label{sec:Lnu}

We can now express a generalized neutrino luminosity that depends on the neutrino production efficiency, $ \kappa_{p\gamma} = t_{\rm diff} t^{-1}_{p\gamma} $, where the dominant parameters are $ L_{\rm x} $ and $ M_\bullet $.
In the following, we express the neutrino luminosity $L_\nu(E_\nu)$ as a function of $\gamma_p$ such that $E_\nu = \alpha \gamma_p m_pc^2$ because of the dependence of $s$ on $\gamma_p$. Recall that we set the minimum proton Lorentz factor contributing to the diffuse flux to $\gamma_p \geq 10^4$, corresponding to a minimum neutrino energy of $E_\nu^{\rm m} \simeq 10^{12}$eV. We emphasize that $\gamma_p\ge10^4$ is the range relevant for neutrino production (where
$\rl \to\lc$ and the \citet{mbarek+26a} slope is robust for $\delta B/B \sim 1$ at the largest scales). We do not extrapolate the $\gamma_p^{-s}$ for $\rl\ll\lc$ as this intermediate-scale
spectrum remains an open, simulation-limited question \citep{mbarek+26a}.

If $\kappa_{p \gamma} < 1$, we write (in $\rm erg\,s^{-1}$)
\begin{equation}\label{eq:Lnu_g}
\begin{split}
L_\nu(\geq E_\nu^{\rm m})
\simeq&
\frac{C(r)}{\sigma_p^{1+r}\zeta^{1-r}}
\left(\frac{10^{7}M_\odot}{M_\bullet}\right)
\left(\frac{L_{\rm x}}{10^{44}\,\rm erg\,s^{-1}}\right)^{1+r/2}
\\[-1pt]
\times&
\left( \frac{\gamma_p}{\sigma_p} \right)^{-s+2-r}
\begin{cases}
\displaystyle
f_{\rm BHe}^{\rm x}\frac{L_{\rm x}}{\epsilon_0},
\\[-1pt]
\qquad
\displaystyle
\quad \,\,\,\, \text{if }\frac{\bar\epsilon_{\rm th}}{2\epsilon_0}
<\gamma_p<
\frac{\bar\epsilon_{\rm th}}{2\epsilon_1}
\\[7pt]
\displaystyle
f_{\rm BHe}^{\rm x+UV}
\left[
\frac{L_{\rm x}}{\epsilon_0}
+\frac{L_{\rm UV}}{(r_{\rm UV}/r_c)^2\epsilon_1}
\right],
\\[-1pt]
\qquad
\displaystyle
\quad \,\,\,\, \text{if }\frac{\bar\epsilon_{\rm th}}{2\epsilon_1}
<\gamma_p
\end{cases}
\end{split}
\end{equation}
\noindent with
$C(r)=\dfrac{\xi\sigma_{\rm eff}L_{x,0}}{8\pi c r_{g,0}}\left[\dfrac{e}{10m_p}\sqrt{\dfrac{2L_{x,0}}{c^{5}}}\right]^{r}\approx (3.8\times10^{-9})(2.6\times10^{9})^{r}\ \mathrm{erg}$,
evaluated at $L_{x,0}=10^{44}\,\rm erg\,s^{-1}$ and $r_{g,0}=GM_0/c^2$ for $M_0=10^{7}\,M_\odot$. Here, $f_{\rm BHe}^{\rm x}$ and $f_{\rm BHe}^{\rm x+UV}$ denote the BHe process efficiency evaluated depending on the photon fields. The corresponding $f_{\rm BHe}$ varies depending on the radiation fields (see Appendix~\ref{app:bh} for details).

Now, if $\kappa_{p\gamma} = 1$, i.e., all protons interact with radiation within one $t_{\rm diff}$, and $L_\nu$ depends mostly on $\lx$,
\begin{equation}\label{eq:Lnu_Lx}
\begin{split}
    L_\nu (E_\nu \geq E_\nu^{\rm m}) \simeq \frac{3 \lx }{4} \frac{f_{\rm BHe}}{\sigma_p} \left( \frac{\gamma_p}{\sigma_p} \right)^{-s + 2}
    \\
    \simeq 8\times 10^{43} \frac{L_{\rm x}}{10^{44}\rm \ergs} \frac{f_{\rm BHe}}{\sigma_p} \left( \frac{\gamma_p}{\sigma_p} \right)^{-s + 2}
\end{split}
\end{equation}

\begin{figure}
	\centering
\includegraphics[width=0.46\textwidth,clip=false,trim= 0 0 0 0]{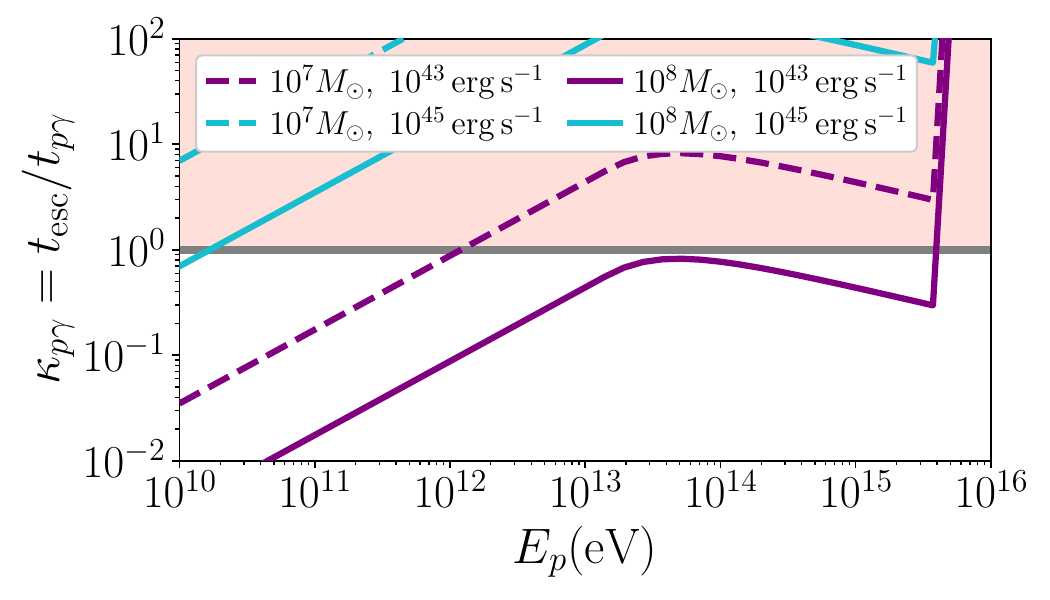}
	\caption{The ratio of escape time $t_{\rm esc} =t_{\rm diff}$ to cooling time $t_{p\gamma}$ as a function of the proton energy $E_p$, based on examples of SMBH masses and X-ray luminosities. The production efficiency is limited by $\kappa_{p\gamma} \leq 1$ as in Eq.~\eqref{eq:Lnu1} to conserve energy. More X-ray luminous AGN are likely to have $\kappa_{p\gamma} = 1$, mostly eliminating the impact of $M_\bullet$ on $L_\nu$.}
	\label{fig:kappa}
\end{figure}

The neutrino production efficiency scales as $\kappa_{p\gamma} \propto \lx^{1 + r/2} M_\bullet^{-1}$. Fig.~\ref{fig:kappa} illustrates the dependence of $\kappa_{p\gamma}$ on both $L_{\rm x}$ and $M_\bullet$. For proton energies above the TeV scale, $\kappa_{p\gamma}$ can approach unity in systems with $M_\bullet \lesssim 10^8 M_\odot$, provided the X-ray luminosity is not too low. More broadly, the efficiency remains well below unity in systems with $M_\bullet \sim 10^9\,M_\odot$ and $\lx \lesssim 10^{44}\,\mathrm{erg\,s^{-1}}$—a regime that is expected to be uncommon, as $\lx$ generally increases with SMBH mass. Consequently, Eq.~\eqref{eq:Lnu_Lx} is expected to hold for most SMBH systems producing the highest-energy neutrinos, with $\lx$ serving as the primary governing parameter.

\section{The Neutrino Diffuse Flux}

We can compute the expected flux of neutrinos $\mathcal{F}(E_\nu) \simeq E_\nu^2 \phi_\nu$
from the convolution over the cosmological distribution of AGN X-ray sources \citep[e.g.,][]{murase+14,ahlers+17,mbarek+23},
\begin{equation}\label{eq:Numain}
	\begin{split}
	 \mathcal{F}(E_\nu) = \frac{c}{4 \pi} \int_0^{z_{\rm max}} \frac{dz}{H(z)} \int_{L_{\rm min}}^{L_{\rm max}} dL_{\rm x} \frac{d\rho}{dL_{\rm x}}(\lx, z) \\ \times L_\nu[\frac{(1+z)E_\nu}{\alpha m_p c^2},L_{\rm x}, \Bar{M}_{\bullet}(z) ]
	\end{split}
\end{equation}
where $\rho(z)$ is the number density of AGN X-ray sources, 
$H(z) = H_0 \sqrt{(1+z)^3 \Omega_M + \Omega_\Lambda}$ is the Hubble parameter, with $\Omega_M \approx 0.3$ and $\Omega_\Lambda \approx 0.7$ for standard $\Lambda$CDM cosmology, and $H_0 = 70$km~s$^{-1}$~Mpc$^{-1}$. The luminosity $L_\nu$ is the source-dependent neutrino luminosity.

We rely on the redshift-dependent modeling of the X-ray luminosity function (XLF) \citep{ueda+14}, necessary for an integration of Eq.~\eqref{eq:Numain}, and include contributions from galaxies with $\lx > 10^{41} \ergs$ (See Appendix~\ref{app:rholx}).
The active SMBH mass function is unlikely to significantly affect the predicted diffuse neutrino flux (see \S\ref{sec:Lnu}). However, we still include it in our modeling to account for potential contributions from sources with large $\mbh$ and low $\lx$. In general, the mass function cannot be directly constrained by the X-ray luminosity function (XLF), since the AGN bolometric luminosity $\lbol$—and therefore $\lx$\footnote{We can relate $\lx$ to $\lbol$ such that $\lx \simeq \lbol/20 $ following the prescription by \citet{marconi+04}.}---depends on both $\mbh$ and the accretion rate, with $\lbol \propto \lambda_E M_\bullet$, where $\lambda_E$ is the Eddington ratio \citep[e.g.,][]{ananna+22}. If an underlying relation between the $M_\bullet$ mass function and XLF exists, it would be obscured by $\lambda_E$. Therefore, in Eq.~\eqref{eq:Numain}, we rely on observations of the most likely masses $\Bar{M}_{\bullet}$ of active SMBHs in different redshift bins. For low redshifts $z< 0.3$, we rely on the BAT AGN survey \citep{ananna+22} to retrieve $\Bar{M}_{\bullet}$, and on the Large Bright Quasar Survey, the Bright Quasar Survey, and the Fall Equatorial Stripe of the Sloan Digital Sky Survey for $0.3 < z\leq 4$ (see Appendix~\ref{app:rholx}). We don't expect our assumptions on $\mbh$ to significantly affect our result since $\kappa_{p\gamma}$ is generally of order unity for neutrino-producing Seyferts (see \S\ref{sec:Lnu}).

\subsection{Neutrino Diffuse Signal} \label{sec:diffNu}

\begin{figure}
	\centering
\includegraphics[width=0.48\textwidth,clip=false,trim= 0 0 0 0]{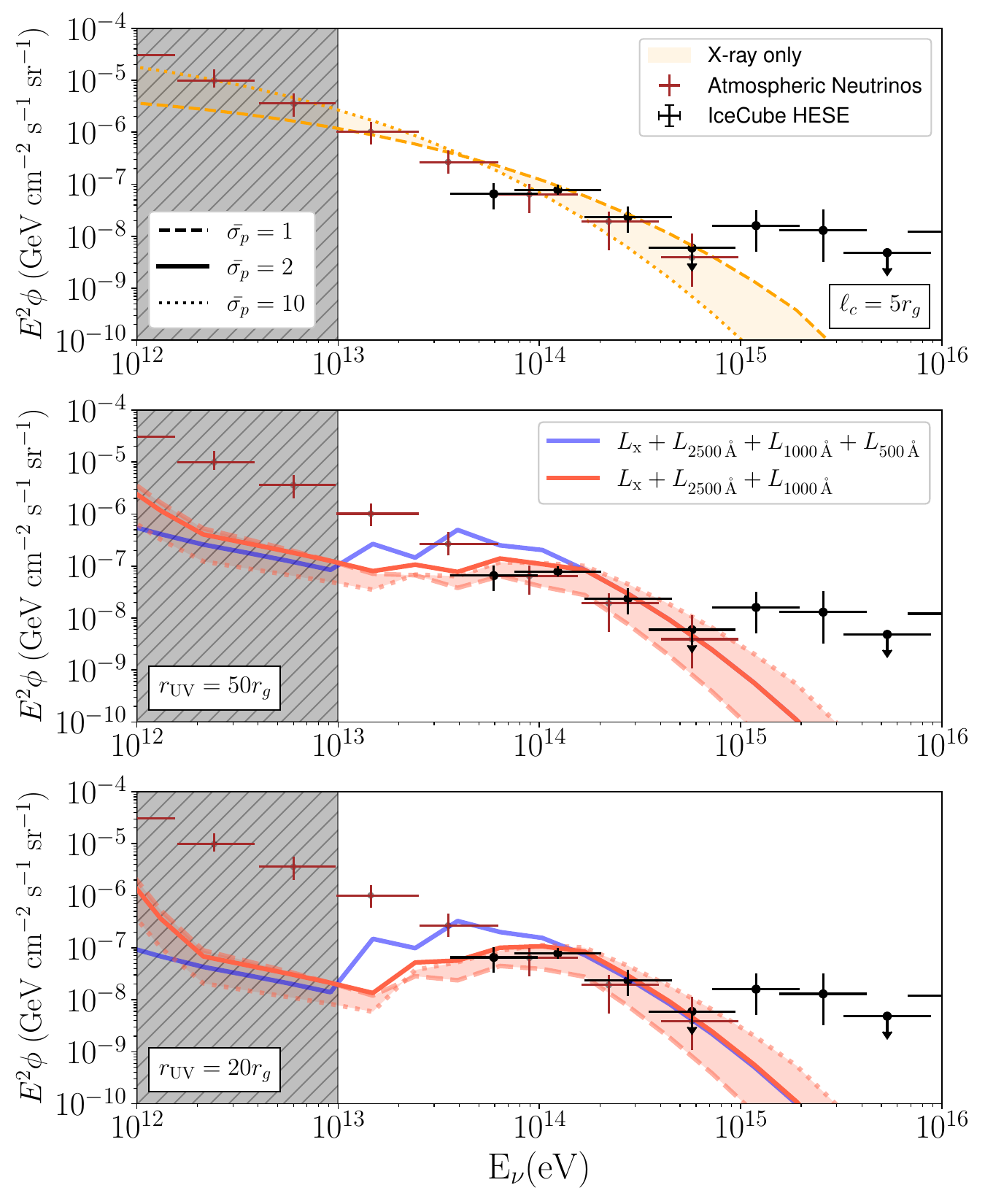}
	\caption{ Expected neutrino flux based on Eq.~\eqref{eq:Numain}, shown for various photon field components near the black hole. Different average proton magnetizations, $\bar{\sigma}_p$, are considered, with a fixed magnetic coherence length $\lc = 5r_g$. Flux values in the dashed region ($E_\nu \leq 10^{13}$eV) reflect an upper bound due to greater uncertainty in the proton distribution at low energies (see \S\ref{sec:p-density} and \S\ref{sec:diffNu}). The results are compared with IceCube’s atmospheric background \citep{ICECUBE15} and extragalactic diffuse HESE flux \citep{ICECUBE20}.}
	\label{fig:diffuse}
\end{figure}

We show in Fig.~\ref{fig:diffuse} the expected  diffuse neutrino flux from coronae of SMBHs depending on the proton magnetization $\sigma_p$, and dominant photon fields in the vicinity of SMBHs. 
We consider a few radiation scenarios,
\paragraph{X-rays only:}
The only considered photons in this case are 2-8~keV X-rays (orange lines in Fig.~\ref{fig:diffuse}). We can see that the neutrino contribution is still significant, suggesting that the X-ray impact should be dominant. This scenario is unlikely as UVs should be abundant in the vicinity of SMBHs, affecting neutrino production.

\paragraph{X-rays and the 2500\AA~component:}
A tight relation between X-rays and the 2500\AA~component is observed in AGNs \citep[e.g.,][]{lusso+10, jin+24}, and is set by the spectral index $\alpha_{\rm ox} = -{\log{(L_{\rm 2keV}/L_{\rm 2500 \text{\AA}})}}/{2.605}$ \citep[e.g.,][]{tananbaum+79,zamorani+81,silverman+05,steffen+06,just+07}. We then obtain a source-dependent $L_{\rm 2500 \text{\AA}}$.
The half-light radius $r_{2500}$ of this component sits at $100 r_g$ for SMBH masses $\gtrsim 10^7 M_{\odot}$ \citep[from microlensing, e.g.,][]{edelson+15}, and could reach $10^3 r_g$ \citep[from reverberation mapping, e.g.,][]{cackett+21}. 
We set $r_{2500} = 500 r_g$ as an average. 

Even a large $2500\,{\rm \AA}$ luminosity does not substantially boost the PeV neutrino output, because the coronal proton spectrum steepens above $E_p\gtrsim10^{15}\,{\rm eV}$. This steepening sets a natural turnover in the diffuse spectrum near $\lesssim1\,{\rm PeV}$: from Eq.~\eqref{eq:slope2}, $s-2\propto \gamma_p^{r}L_{\rm x}^{-r/2}$, implying $\Delta s\gtrsim0.5$ over $\gamma_p=10^4$--$10^7$ even for luminous coronae with $L_{\rm x}\gtrsim10^{45}\,{\rm erg\,s^{-1}}$. Bethe--Heitler cooling at $10^{14}$--$10^{16}\,{\rm eV}$ further sharpens this break.

PeV neutrinos are therefore unlikely to arise mainly from turbulent proton acceleration inside the X-ray corona. Instead, a PeV component likely requires re-acceleration outside the corona (Section~\ref{sec:cr-outflow}), for example if a small fraction of coronal protons is injected into a black-hole-driven outflow and re-accelerated before interacting with the $2500\,{\rm \AA}$ radiation field, whose characteristic photon energy is $\epsilon_{2500}\simeq 5\,{\rm eV}$. 
Because this putative reacceleration occurs at $\gg r_g$, outside the radiatively compact corona, its accompanying $\gamma$-rays may escape absorption. There, the 2500~\AA{} continuum can provide the $p\gamma$ target, selecting PeV neutrinos: $\gamma_p\gtrsim\bar\epsilon_{\rm th}/(2\epsilon_{2500})\simeq1.5\times10^7$ implies $E_\nu\gtrsim0.7$~PeV, whereas in-situ X-ray targets produce lower-energy neutrinos. This speculative, low-power outflow component does not affect our sub-PeV results.

\paragraph{Adding a UV continuum component}

Additional contributions from higher-energy photons ($\epsilon>\epsilon_{2500}$) are a ubiquitous feature of AGN spectra \citep[e.g.,][]{zheng+97,telfer+02}, typically arising from the disk continuum. The disk origin of the broad optical/UV continuum is also supported by its strong wavelength-to-wavelength correlation \citep{jin+24}. This emission is produced in the inner regions of the accretion disk on sub-light-day scales ($\gtrsim50$--$100\,r_g$) in the case of, e.g., NGC~5548 \citep[e.g.,][]{edelson+15}. We assume that this UV continuum, which dominates the local radiation field, is consistently produced within tens of $r_g$ across a broad range of SMBH masses.

Observations suggest that the luminosity near 900--1400\AA\ often exceeds that at both 1400\AA\ and 2500\AA, with typical ratios of order $\sim3$ relative to $L_{2500}$ \citep[e.g.,][]{zheng+97,telfer+02,shull+12,stevans+14}. The X-ray/UV luminosity correlation remains similar to the 2500\AA\ relation down to at least 1280\AA\ \citep{jin+24}. At shorter far-UV wavelengths, however, an analogous relation remains poorly established, although this component likely contributes to the radiation field near SMBHs \citep{edelson+15}.

Motivated by these results, we include two far-UV components: $L_{1000\mathring{\mathrm{A}}}=3L_{2500\mathring{\mathrm{A}}}$ and $L_{500\mathring{\mathrm{A}}}=L_{1000\mathring{\mathrm{A}}}$ \citep[e.g.,][]{shull+12}. Both components are initially assumed to originate within $r_{\rm UV}=50r_g$, with $r_{\rm UV}=20r_g$ subsequently considered as a more compact case. Other components remain uncertain because Galactic absorption strongly suppresses direct extreme-UV measurements, although some studies have argued for their presence \citep[e.g.,][]{malkan+82,bechtold+87,mathews+87}. Our conclusions depend mainly on the photon energy density in these wavelength bands rather than on the detailed spectral shape.

With increasing IceCube exposure and the development of next-generation neutrino detectors \citep[e.g., KM3NeT;][]{martinez+16}, sub–100 TeV extragalactic neutrino detections may offer an indirect integrated view of these otherwise obscured extreme-UV fields. So far, recent IceCube results have enhanced our understanding of the sub–100 TeV astrophysical neutrino sky \citep{ICECUBE24}, helping to better constrain the extragalactic contribution.

\paragraph{Synthesis}
Fig.~\ref{fig:diffuse} presents the predicted neutrino flux based on Eq.~\eqref{eq:Numain} for energies $E_\nu \geq 10^{12}$eV (corresponding to $\gamma_p \geq 10^4$). The flux at lower energies ($E_\nu \leq 10^{13}\,\mathrm{eV}$) should be regarded as a normalizing upper bound and is likely at much lower levels \citep[see][for more details]{mbarek+26a}. The particle spectrum at these lower energies depends on small-scale interactions with $\rl \ll \lc$, which remain the subject of active research \citep[e.g.,][]{boldyrev+26}. This choice of $\gamma_p \geq 10^4$ ensures that the proton Larmor radius $\rl$ approaches the turbulence coherence length for most of the considered sources (see \S\ref{sec:p-density}).

The upper panel shows the flux obtained under the limiting assumption that only X-rays interact with protons near black holes. This case gives a reasonable match to the flux around $100\,{\rm TeV}$, but overestimates the extragalactic diffuse component at lower energies. In the middle and lower panels, we include contributions from UV photon fields. These photons produce noticeable suppression below $E_\nu\sim10^{13}\,{\rm eV}$ through BHe interactions, which deplete the proton population before efficient neutrino production occurs. Above $E_\nu\gtrsim10^{13}\,{\rm eV}$, however, the same UV fields enhance the neutrino flux once the threshold for $p\gamma$ interactions is reached. With $\lc=5r_g$ and $\bar{\sigma}_p\approx1$ for all sources, we find a good fit to the IceCube data, with the detailed shape depending on the adopted half-light radius of the far-UV components.

Including the $500\,{\rm \AA}$ component primarily enhances the diffuse flux above
$E_\nu \simeq \alpha\,\bar{\epsilon}_{\rm th}\,m_pc^2/(2\epsilon_{500}) \sim 100\,{\rm TeV}$,
where $\gamma_p$ crosses the $p\gamma$ threshold in Eq.~\eqref{eq:tpg}. In this regime, the additional
$L_{\rm UV}/[(r_{\rm UV}/r_c)^2\epsilon_1]$ term in Eq.~\eqref{eq:Lnu_g} dominates over the accompanying reduction in $f_{\rm BHe}$, because the relevant per-photon ratio is
$\xi\sigma_{\rm eff}/(\hat{\xi}_{\rm BHe}\hat{\sigma}_{\rm BHe})\approx 60$. At lower $E_\nu$, the $500\,{\rm \AA}$ photons lie above the BHe threshold but below the $p\gamma$ threshold. They therefore contribute only to $t_{\rm BHe}^{-1}$ and act purely to suppress the proton population.

Although the results in Fig.~\ref{fig:diffuse} rely on simplified assumptions about the average $\lc$, the average $\bar{\sigma}_p$, and the dominant photon fields near SMBHs, they reveal a simple scaling that is broadly consistent with current sub-PeV neutrino observations, including the decrease in flux below $\sim10^{15}\,{\rm eV}$. Overall, we find that our model favors an average magnetization of order $\bar{\sigma}_p\approx1$, although this value should be interpreted as tracing the most powerful sources that dominate the neutrino flux. With the continued operation of IceCube and the advent of KM3NeT, these results may provide a promising way to probe obscured photon fields in the extreme-UV environments around SMBHs.

\subsection{Corresponding Diffuse $\gamma$-rays}

Photomeson interactions also produce high-energy $\gamma$-rays, but these photons are unlikely to escape compact coronae directly. They instead undergo Breit--Wheeler pair production on ambient photons of energy $\sim m_e^2c^4/\epsilon_\gamma$, triggering electromagnetic cascades that reprocess the injected power to lower energies. In NGC~1068, for example, GeV--TeV coronal emission must be suppressed to avoid exceeding the emission associated with star formation, suggesting that $\gtrsim{\rm TeV}$ coronal $\gamma$-rays may emerge instead as cascaded MeV emission \citep[e.g.,][]{ajello+23}.

The MeV band is therefore a key counterpart channel. Legacy measurements were sensitivity-limited \citep[e.g.,][]{COMPTEL96,EGRET98}, leaving the MeV background comparatively unexplored \citep{meyer+19}. While GeV--TeV emission may be present in star-forming or active galaxies, it can originate in extended regions that are inefficient neutrino sources compared with compact coronae such as NGC~1068 \citep{padovani+24}. The more direct test is whether hidden hadronic power produces a diffuse, cascaded MeV component. The possible similarity between the TeV diffuse neutrino flux and the diffuse MeV $\gamma$-ray flux inferred from legacy data \citep{COMPTEL96,EGRET98} makes this connection especially timely.

COSI, a NASA Small Explorer Compton telescope planned to survey the sky over $\simeq0.2$--$5\,{\rm MeV}$, will directly probe this cascade window with wide-field imaging, spectroscopy, and polarization capabilities \citep{COSI24}. Together with future MeV concepts such as AMEGO-X, COSI motivates detailed cascade calculations that connect hidden coronal neutrino sources to testable diffuse and point-source MeV $\gamma$-ray signatures \citep{AMEGOX22,murase+24}.

\section{Local Source Expectations and the Dominant Redshift of the Diffuse Flux}
\label{sec:local}

The diffuse neutrino flux is weighted by the product of $L_\nu$ and the co-moving source density $d\rho/dL_{\rm x}$, 
integrated over $z$ as in Eq.~\eqref{eq:Numain}. Since 
$L_\nu \propto L_{\rm x}^{2.15} M_\bullet^{-1}$ (Eq.~\eqref{eq:Lnu_g}) 
for $\kappa_{p\gamma}<1$, and the XLF \citep{ueda+14} peaks near $z \sim 1$--$2$ for sources with $L_{\rm x} \gtrsim 10^{44}$~erg~s$^{-1}$, the bulk of the predicted diffuse flux originates at $z \sim 0.5$--$2$. At these redshifts, the co-moving density of luminous AGN is $\sim 10$--$100\times$ higher than locally \citep{ueda+14}, and the characteristic SMBH mass is $\Bar{M}_{\bullet} \gtrsim 10^{8}\,M_\odot$ (see \S\ref{sec:Lnu}). The efficiency $\kappa_{p\gamma} \propto L_{\rm x}^{1.15} M_\bullet^{-1}$ approaches or exceeds unity for such systems provided $L_{\rm x} \gtrsim 10^{44}$--$10^{45}$~erg~s$^{-1}$, meaning the most luminous AGN at cosmic noon contribute disproportionately to the diffuse flux despite the $\sigma_p$ uncertainty. Conversely, at $z \lesssim 0.1$ the AGN number density is suppressed by a factor of $\sim 10^2$ relative to $z \sim 1$ \citep{ueda+14}, and typical X-ray luminosities of local Seyfert galaxies fall in the range 
$L_{\rm x} \sim 10^{42}$--$10^{43.5}$~erg~s$^{-1}$ \citep{ricci+17,koss+17}, yielding softer proton spectra (Eq.~\eqref{eq:slope2}) and lower $\kappa_{p\gamma}$. The local universe is then expected to be a subdominant contributor to the \emph{diffuse} flux, though individual nearby sources can produce detectable point-source signals because of the $d^{-2}$ flux enhancement.

This $z\sim0.5$--$2$ weighting ties the diffuse neutrino sky to the cosmic X-ray background (XRB). Above $\sim1$~keV, and especially in the $2$--$10$~keV band, the XRB is dominated by integrated AGN emission, with much of the contribution coming from similar redshifts \citep{brandt+05,gilli+07,ueda+14,aird+15}. The two backgrounds may therefore trace part of the same projected large-scale structure. On very large angular scales, $\sim100^\circ$, the XRB is nearly isotropic, with only weak dipole-level anisotropy. The anisotropy most directly associated with the cosmological AGN population is instead expected mainly on arcminute-to-degree scales \citep{scharf+00}. Such structure is strongly smoothed in cascade-dominated neutrino samples, particularly at $E_\nu\sim10$--$300\,{\rm TeV}$ where the event statistics are largest but angular uncertainties are broad. Track-like events provide the more suitable channel, since their sub-degree to degree-scale angular resolution is comparable to the expected clustering scale. At PeV energies, reconstruction improves but the statistics become sparse. 
Since the present model primarily predicts a sub-PeV neutrino component, the most direct test is a correlated anisotropy in the $10$--$300\,{\rm TeV}$ band, where event statistics are largest, while any extension to PeV energies would probe whether proton re-acceleration near black holes remains efficient at the highest energies. 
KM3NeT will extend the same test to the Southern sky. By contrast, a $\sim100^\circ$-scale neutrino modulation would not be limited by angular resolution, but would primarily probe dipole-level structure rather than the clustering signal expected from the dominant $z\sim0.5$--$2$ AGN population.

\paragraph{NGC~1068 is a special case.}
Among local AGN, NGC~1068 
($d \approx 10$~Mpc, $L_{\rm x} \approx 4\times10^{43}$~erg~s$^{-1}$, 
$M_\bullet \approx 10^7\,M_\odot$) sits near the optimum of every parameter 
entering $L_\nu$ \citep{ricci+17,ananna+22,IceCube-NGC1068}. 
The proton energy density normalization 
$\propto L_{\rm x}/(\sigma_p M_\bullet^2)$ is favorable because $M_\bullet$ is modest, and $\kappa_{p\gamma} \sim 1$ is achieved at TeV proton energies for this mass and luminosity, placing it in the saturated regime $L_\nu \propto L_{\rm x}/\sigma_p$ (Eq.~\eqref{eq:Lnu_Lx}). 
For NGC~1068, adopting $\sigma_p\sim1$ and $\zeta\simeq0.5$ \citep{mbarek+24} yields an effective proton index $s_{\rm eff}\simeq3$ at the energies relevant to the IceCube signal. 

\paragraph{IceCube AGN candidates.}
The additional local AGN with IceCube hints---NGC~4151, Circinus, NGC~4945, 
and NGC~7469 \citep{ICECUBE_25a,ICECUBE_25b}---are consistent with the model within the uncertainties on $\sigma_p$ and $\zeta$. NGC~4151 ($d = 15.8 \pm 0.4$~Mpc, $L_{\rm x} \approx 10^{43}$~erg~s$^{-1}$, $M_\bullet \approx 3\times10^7\,M_\odot$ \citep{ricci+17,yuan+20,roberts+21,bentz+22}) is the most direct NGC~1068 analogue in the Northern sky. Its larger $M_\bullet$ pushes $\kappa_{p\gamma}$ below unity at TeV energies, but a moderate $\sigma_p$ could harden the slope sufficiently to compensate in the 
$10$--$100$~TeV band. Circinus ($d \approx 4.2$~Mpc, $L_{\rm x} \approx 5\times10^{42}$~erg~s$^{-1}$, $M_\bullet \approx 1.5\times10^6\,M_\odot$) benefits from proximity and a tiny corona ($r_c \propto M_\bullet$), giving $\kappa_{p\gamma} \sim 1$ 
despite low $L_{\rm x}$. Its main liability is a soft intrinsic spectrum from 
Eq.~\eqref{eq:slope2} \citep{greenhill+03,ricci+17}. 
NGC~4945 ($d \approx 3.7$~Mpc, $L_{\rm x} \approx 10^{43.2}$~erg~s$^{-1}$) has the most 
favorable combination of $L_{\rm x}/d^2$ and $L_{\rm x}/M_\bullet^2$ of the 
entire sample, and is likely the strongest Southern-sky target for KM3NeT \footnote{The heavy Compton-thick column 
($N_H \sim 4\times10^{24}$~cm$^{-2}$) introduces a factor of $\sim 2$--$3$ 
uncertainty in the intrinsic $L_{\rm x}$ \citep{ricci+17,ananna+22}}. 
For NGC~7469 ($d \approx 65$~Mpc, $L_{\rm x} \approx 10^{43.2}$~erg~s$^{-1}$, 
$M_\bullet \approx 1.2\times10^7\,M_\odot$), the two candidate IceCube 
neutrinos at $E_\nu \sim 100$--$200$~TeV with no lower-energy excess would be a signature of elevated $\sigma_p$. 

\section{Potential for Cosmic-Ray-Driven Outflows}\label{sec:cr-outflow}

The neutrino flux from NGC~1068 requires a substantial coronal proton luminosity $L_p$ \citep[e.g.,][]{mbarek+24}, and analogous Seyferts should host comparable populations. The in-situ neutrino-producing population is governed by diffusion, but diffusion is too slow to provide efficient bulk transport (see \S\ref{sec:escape} and Fig.~5 in \citet{mbarek+26a} for NGC~1068). The corresponding outward flux is suppressed by $t_{\rm lc}/t_{\rm diff}\ll1$, implying that most of the proton energy is radiated locally. A small residual fraction may nevertheless avoid local losses, either by streaming along the dominantly poloidal field expected near accreting black holes or by being advected with magnetized coronal plasma. These escaping cosmic rays (CRs) could contribute to a CR-loaded or CR-driven outflow and may even provide seeds for larger-scale winds \citep[e.g.,][]{ruszkowski+17,chan+19,hopkins+20}. Realizing a dynamically important outflow through this channel, however, requires a sufficiently large escaping CR power.

We parametrize this leakage by $\eta_{\rm cr}$,
\begin{equation}\label{eq:eta-def}
L_{\rm cr}(\gamma_{\rm cr}) = \eta_{\rm cr}L_p(\gamma_{\rm cr}),
\qquad
L_p = 4\pi r_c^2 c\,E_p^2\frac{dn_p}{dE_p},
\end{equation}
where $\gamma_{\rm cr}$ is the Lorentz factor in the escaping channel and $L_p$ is the luminosity of the diffusively confined population evaluated at $\gamma_p=\gamma_{\rm cr}$. Since $L_p\simeq U_p/t_{\rm lc}$, with $U_p=4\pi r_c^3 E_p^2dn_p/dE_p$ the nonthermal proton energy, $\eta_{\rm cr}$ is equivalently the fraction of $U_p$ removed per light-crossing time. In steady state, the escape-channel loss rate cannot exceed the diffusive supply rate, so $\eta_{\rm cr}\lesssim t_{\rm lc}/t_{\rm diff} \sim 3\times10^{-3}\text{--}10^{-2}$.
Explicitly, $\dot U_{p,\rm cr}\simeq \eta_{\rm cr}U_p/t_{\rm lc}$ must satisfy $\dot U_{p,\rm cr}\lesssim \dot U_{p,\rm diff}\simeq U_p/t_{\rm diff}$, which directly gives $\eta_{\rm cr}\lesssim t_{\rm lc}/t_{\rm diff}$.
Thus $\eta_{\rm cr}$ should be interpreted as a ceiling on the escaping-power fraction, rather than as a generic escape efficiency.

Combining Eq.~\eqref{eq:pden1} with Eq.~\eqref{eq:eta-def} gives
\begin{equation}\label{eq:lcr}
L_{\rm cr} \lesssim
\frac{2\,\eta_{\rm cr}L_{\rm x}}{\sigma_p}
\left(\frac{\gamma_p}{\sigma_p}\right)^{-s+2},
\end{equation}
and saturating the bound $\eta_{\rm cr} \lesssim t_{\rm lc}/t_{\rm diff}$ yields (see Eq.~\eqref{eq:tesc}):
\begin{equation}\label{eq:lcr_bound}
L_{\rm cr}\lesssim 
\frac{5\times10^{41}{\rm erg\,s^{-1}}}{\sigma_p}
\left(\frac{L_{\rm x}}{10^{44}{\rm erg\,s^{-1}}}\right)^{0.85}
\left(\frac{\gamma_p}{10^4}\right)^{0.3}
\left(\frac{\gamma_p}{\sigma_p}\right)^{-s+2}
\end{equation}
If $\eta_{\rm cr}$ reaches the maximal value allowed by $t_{\rm lc}/t_{\rm diff}$, the escaping CR power can be substantial, especially as $s \to 2$. It can approach the canonical CR power supplied by Galactic supernovae, $\sim10^{41}\,{\rm erg\,s^{-1}}$ \citep[e.g.,][]{blandford+87}.

Such an escaping population could in principle be re-accelerated in shocked environments in the vicinity of black holes. This possibility is motivated by the broader CR-wind literature, where CR streaming can excite instabilities that steepen into shocks and strongly modify the outflow structure. CR-streaming-driven winds could develop strong shocks, producing a staircase-like CR pressure profile and multiphase gas \citep[e.g.,][]{quataert+22a,quataert+22b}. Analogous shocked coronal or nuclear outflows could therefore provide sites where leaked coronal CRs are re-energized before escaping or interacting.

Separately, the same escaping population could be injected into AGN jets, where further acceleration to ultra-high energies is possible \citep{caprioli15,mbarek+19,mbarek+21a,mbarek+25a}. Because these particles originate near the disk, they may include a non-negligible heavy-element component, consistent with prominent Fe K$\alpha$ reflection features near black holes \citep[e.g.,][]{wilkins+22,laha+25} and inferred disk Fe abundances that can be comparable to solar values \citep{tomsick+18,huang+23}.

\section{Conclusions}
We have outlined a framework that connects observable properties of SMBH coronae to a generalized neutrino luminosity. The source-dependent luminosity function in Eq.~\eqref{eq:Lnu_g}, built on the magnetized-turbulence scaling for nonthermal spectra \citep{mbarek+26a}, can naturally reproduce the diffuse sub-PeV neutrino flux observed by IceCube in Eq.~\eqref{eq:Numain}, and ties the physical conditions of AGN coronae to high-energy neutrino emission. Our principal findings are:

\begin{itemize}
    \item The coronal neutrino luminosity is controlled mainly by $L_{\rm x}$ and $\sigma_p$, with only weak dependence on $M_\bullet$. After cosmological integration, this population reproduces the sub-PeV diffuse IceCube flux, but the steepening of the proton spectrum prevents the most luminous sources from extending the emission to PeV energies. A genuine PeV component likely requires re-acceleration in a CR-loaded outflow (\S~\ref{sec:cr-outflow}). Since the diffuse flux is dominated by AGN at $z\sim0.5$--$2$, the same redshifts that dominate the $2$--$10$~keV XRB, the model is testable via a degree-scale cross-correlation between IceCube tracks and the XRB, with KM3NeT extending the test to the Southern sky.
    \item Reasonable variations in $\sigma_p$, the coherence length $\lc$, and the coronal size $r_c$ shift the predicted flux by factors of a few, but do not qualitatively alter the agreement with observations. A simplified treatment of the AGN UV photon field suffices to capture the dominant features of neutrino production.
    \item Unlike \citet{murase+20b}, where slow stochastic acceleration lets Bethe-Heitler losses terminate the proton spectrum at $\sim0.1$--$1$~PeV, the faster acceleration in moderately magnetized plasma adopted here \citep{mbarek+26a} keeps $t_{\rm acc}\ll t_{\rm BHe}$. Bethe-Heitler losses therefore could suppress the proton population between $10^{14}$ and $10^{16}$~eV, imprinting a spectral feature rather than a hard cutoff.
    \item A small but non-negligible fraction of accelerated protons can stream out along the dominantly poloidal coronal field, powering a CR-driven outflow (Eq.~\eqref{eq:lcr}). The escape fraction $\eta_{\rm cr}\lesssim t_{\rm lc}/t_{\rm diff}$ depends mostly on $L_{\rm x}$. Saturating this bound yields a per-source CR power that can be substantial. Quantifying $\eta_{\rm cr}$ with global simulations is the next step, and the same machinery extends naturally to stellar-mass black holes \citep{fang+24}.
    \item The predicted neutrino spectrum near $\sim 10$--$100$~TeV is sensitive to the extreme-UV photon field, a band that is largely inaccessible to direct observation due to Galactic interstellar absorption. With increasing IceCube exposure and the advent of KM3NeT, the diffuse sub-100~TeV flux---together with spectra of resolved Seyferts---could serve as an indirect, integrated probe of these obscured fields.
    \item The predicted diffuse flux should statistically trace the extragalactic X-ray background, and is not inconsistent with the reported correlations between the diffuse neutrino sky and other radiatively-efficient AGN tracers, which themselves correlate with $L_{\rm x}$ in a galaxy-type-dependent way.
    \item A more refined source-by-source integration that measures $L_{\rm x}$ and $M_\bullet$ independently, and accounts for enhanced escape in regions of strong guide field, will sharpen both the diffuse prediction and the implied outflow energetics.
\end{itemize}

The dominant uncertainties in this scenario are the proton magnetization $\sigma_p$, the magnetic coherence length $\lc$, and the structure of the turbulence on scales $\rl \ll \lc$. Reducing these uncertainties is intrinsically a multi-community problem. Anchoring $\sigma_p$ requires combining global GRMHD simulations of accretion flows, which set the magnetic-energy budget and the location of the corona, with kinetic (PIC) studies that translate those conditions into a self-consistent thermal-plus-nonthermal proton population, and with X-ray observations that constrain the coronal compactness and optical depth and thereby the magnetization itself. 
Finally, the proton distribution at $\rl \ll \lc$, which controls the low-energy tail of the predicted neutrino spectrum, connects directly to ongoing work in the broader turbulence community \citep[e.g.,][]{mbarek+26a,boldyrev+26}. Progress along these axes will determine whether the sub-PeV neutrino sky becomes a quantitative probe of plasma conditions in the vicinity of SMBHs.

\balance
\begin{acknowledgments}
R.M. is supported by a Lyman Spitzer Fellowship at Princeton University. I would like to thank Kohta Murase and Sasha Philippov for useful conversations. 
\end{acknowledgments}

\bibliography{Total,Total_cont}

\balance
\appendix

\section{Energy balance in a turbulent Comptonizing corona}
\label{app:uxub}

We summarize the energy-budget argument from \citet{groselj+24} underlying the relation between the X-ray radiation energy density and the turbulent magnetic energy density. Let the turbulent magnetic energy density be $U_{\rm B}$ and let it dissipate on an eddy time $t_0$, so that the volumetric heating rate is $\dot{u}_{\rm heat}\sim U_{\rm B}/t_0$. Seed photons with number density $n_{\rm seed}$ and characteristic energy $\epsilon_0$ have energy density $U_{\rm seed}=n_{\rm seed}\epsilon_0$. After Comptonization, they escape with characteristic energy $\epsilon_{\rm out}=A\epsilon_0$, where $A$ is the Compton amplification factor, and therefore gain an energy density
\begin{equation}
\Delta U_\gamma \simeq n_{\rm seed}(\epsilon_{\rm out}-\epsilon_0)
= U_{\rm seed}(A-1).
\end{equation}
If the system has reached steady state, the rate at which photons gain energy from the plasma must equal the rate at which that gained energy leaves the system as escaping radiation. Denoting the photon residence time by $t_{\gamma,{\rm esc}}$:
\begin{equation}
\frac{U_{\rm seed}(A-1)}{t_{\gamma,{\rm esc}}}
\sim
\frac{U_{\rm B}}{t_0},
\qquad
A-1\sim
\frac{U_{\rm B}}{U_{\rm seed}}\frac{t_{\gamma,{\rm esc}}}{t_0}.
\label{eq:app_amp}
\end{equation}
Thus the seed photon field affects the cooling, temperature, spectral slope, and amplification factor, but at fixed turbulent heating a larger $U_{\rm seed}$ mainly lowers $A$: the same dissipated power is shared among more photons.

\begin{figure}
	\centering
\includegraphics[width=0.46\textwidth,clip=false,trim= 0 0 0 0]{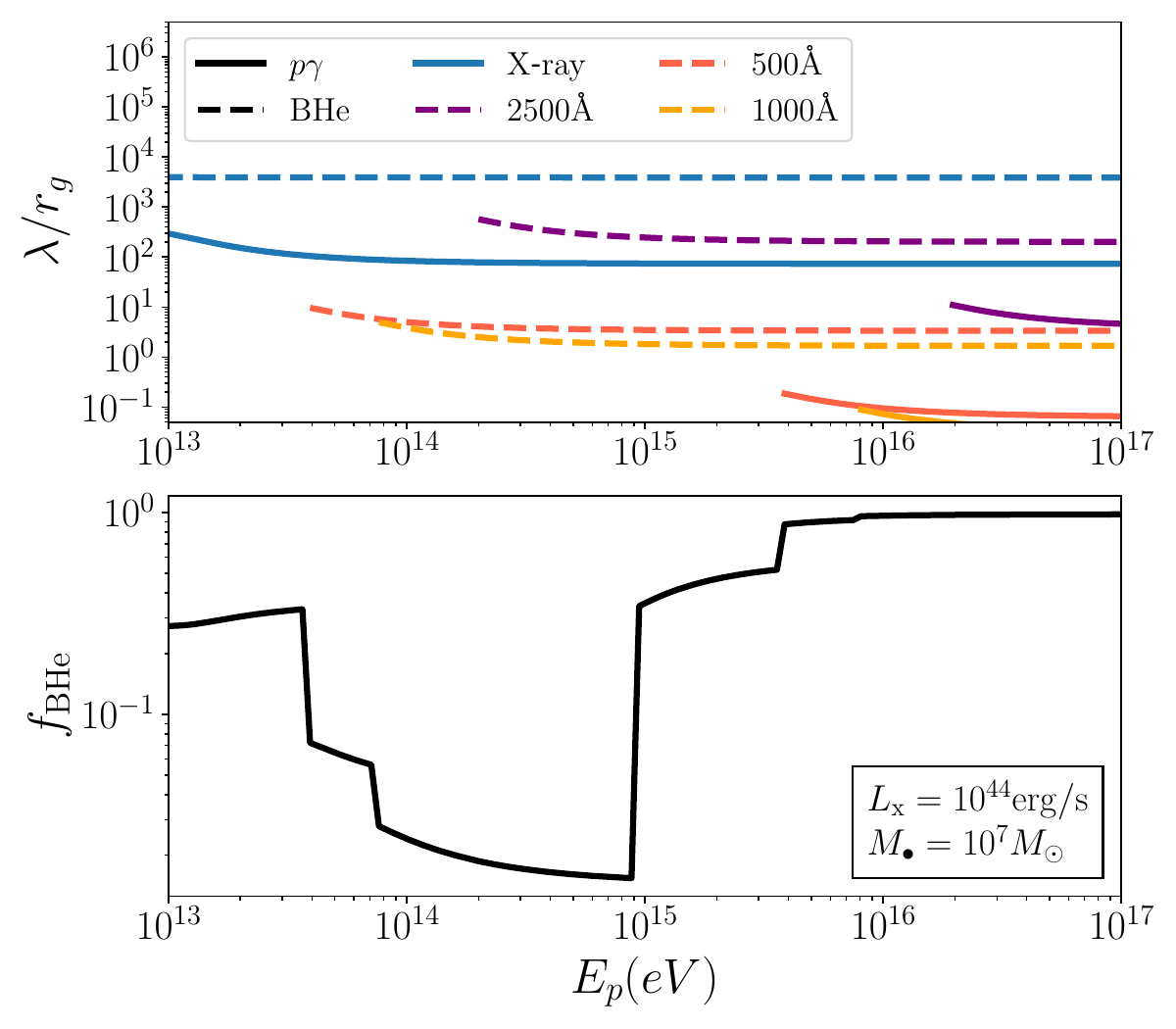}
	\caption{Upper Panel: Cooling time for photomeson ($p\gamma$) and Bethe-Heitler (BHe) for an example galaxy with $\lx = 10^{44} \ergs$ and $M_\bullet = 10^7 M_{\odot}$ for the dominant photon field components considered in this study.
	Lower Panel: fraction of protons available for $p\gamma$ interactions due to BHe suppression for the considered photon fields in the upper panel. A significant suppression is expected for protons with energies ranging from 10 TeV to a few PeV.}
	\label{fig:tpgtbh}
\end{figure}

The Comptonized radiation energy density is
\begin{equation}
U_{\rm x} \sim n_{\rm seed}\epsilon_{\rm out}
= A U_{\rm seed}
\simeq U_{\rm seed}+U_{\rm B}\frac{t_{\gamma,{\rm esc}}}{t_0}.
\label{eq:app_ux}
\end{equation}
When the Comptonized component dominates over the injected seed energy, $U_{\rm seed}\ll U_{\rm B}(t_{\gamma,{\rm esc}}/t_0)$, Eq.~\eqref{eq:app_ux} reduces to
\begin{equation}
U_{\rm x}\sim U_{\rm B}\frac{t_{\gamma,{\rm esc}}}{t_0}.
\label{eq:app_uxub}
\end{equation}
Therefore $U_{\rm x}\sim U_{\rm B}$ follows when photons escape on approximately an eddy time, $t_{\gamma,{\rm esc}}\sim t_0$, while $U_{\rm x}>U_{\rm B}$ is possible when photons are retained for longer than an eddy time. This means that magnetic/turbulent energy is transferred to the radiation field, and $U_{\rm x}$ measures the Comptonized photon energy stored before escape. 

The total X-ray power is controlled primarily by the turbulent heating rate, whereas the seed photon field controls how that power is distributed among photons through Eq.~\eqref{eq:app_amp}. More seed photons give lower amplification and a softer spectrum; fewer seed photons give higher amplification and a harder spectrum, until cooling inefficiency or pair balance modifies the steady state.

\section{Bethe-Heitler interactions}\label{app:bh}
We can compute the Bethe-Heitler (BHe) cooling time in a similar fashion to $p\gamma$ interactions. First, we estimate the product of the inelasticity and an effective cross section $\xi_{\rm BHe} \sigma_{\rm BHe}$ for BHe interactions. 
From Eq.~8 and~9 in \citet{blumenthal+70}, \citet{chodorowski+92} derive Eq~3.1, which leads to the results presented in Fig.~2 of their work. 
We can see that $\xi_{\rm BHe} \sigma_{\rm BHe}$ peaks at $\simeq 2.5 r_0^2 \alpha_s \frac{Z^2 m_e}{A m_p}$ for $20 \bar{\epsilon}_{\rm BHe}$, where $\bar{\epsilon}_{\rm BHe} = 2$MeV is the BHe energy threshold in the proton frame, $Z$ the charge of the particle, $A$ its atomic mass, $\alpha_s = 1/137$ is the fine structure constant, and $r_0 = \sqrt{\frac{3 \sigma_T}{8 \pi}}$ is the classical electron radius with $\sigma_T$ the Thomson cross section.
This yields an effective value of $\hat{\xi}_{\rm BHe} \hat{\sigma}_{\rm BHe} \sim 7.5  \times 10^{-31} $cm$^{2}$ \citep[consistent with][]{murase+10b}. Upon including the full energy-dependent description of 
$\xi_{\rm BHe} \sigma_{\rm BHe}$, no significant differences were noted.

The Bethe-Heitler (BHe) cooling time becomes,
\begin{equation}\label{eq:BHe}
	t^{-1}_{\rm BHe}(\gamma_p) \approx \frac{c}{2 \gamma_p^2} \int^{\infty}_{\bar{\epsilon}_{\rm BHe}/2\gamma_p} d\epsilon \frac{dn_{\rm x}}{d\epsilon} \epsilon^{-2}  \int^{2 \gamma_p \epsilon}_{\bar{\epsilon}_{\rm BHe}}  \epsilon ' \xi_{\rm BHe} \sigma_{\rm BHe}(\epsilon ') d\epsilon '
\end{equation}
Using $\hat{\xi}_{\rm BHe} \hat{\sigma}_{\rm BHe}$ can simplify the integration, but a fit to the $\xi_{\rm BHe} \sigma_{\rm BHe}$ function \citep[Fig.~2 in][]{chodorowski+92}, can also be used to have a more accurate---albeit quite similar---result.
Bethe-Heitler interactions can have a significant impact on the available population of protons because they have a lower interaction threshold than photomeson interactions.

In Fig.~\ref{fig:tpgtbh}, we present the proton suppression due to BHe processes, as explained in the text, for the example of a galaxy with $\lx = 10^{44}\ergs$ and $\mbh = 10^7 M_{\odot}$. 
Proton suppression is mainly due to the relative importance of $p\gamma$ and BHe interactions. Note that $ \epsilon_{\rm max} \lesssim 500$~keV generally represents the highest energy range for X-rays from AGN coronae and, consequently, defines the lowest particle energy band influenced by $ p\gamma $ interactions with these X-rays.

\section{X-ray Luminosity Function}\label{app:rholx}

The term $\frac{d\rho}{d\lx}(\lx, z)$ is the $z$-dependent X-ray luminosity function of the AGN sources per luminosity per comoving volume. The luminosity function can be rewritten as $\frac{d\rho}{d\lx}(\lx, z) = \frac{d\rho}{d\lx}(\lx, z=0) \Xi(z, \lx)$, where $\frac{d\rho}{d\lx}(\lx, z=0)$ is modeled by two connected power laws at the break luminosity $10^{43.97}$\ergs, 
and $\Xi(z, \lx)$ is a luminosity-dependent cosmological evolution factor 
\citep{ueda+14}. 
The luminosity function for different redshifts is plotted in the upper panel of Fig~\ref{fig:rho} for $\lx \geq 10^{41} \ergs$ based on modeling by \citet{ueda+14}.

\section{Distribution of active supermassive black holes}\label{app:rhoMbh}

\begin{figure}
	\centering
\includegraphics[width=0.46\textwidth,clip=false,trim= 0 0 0 0]{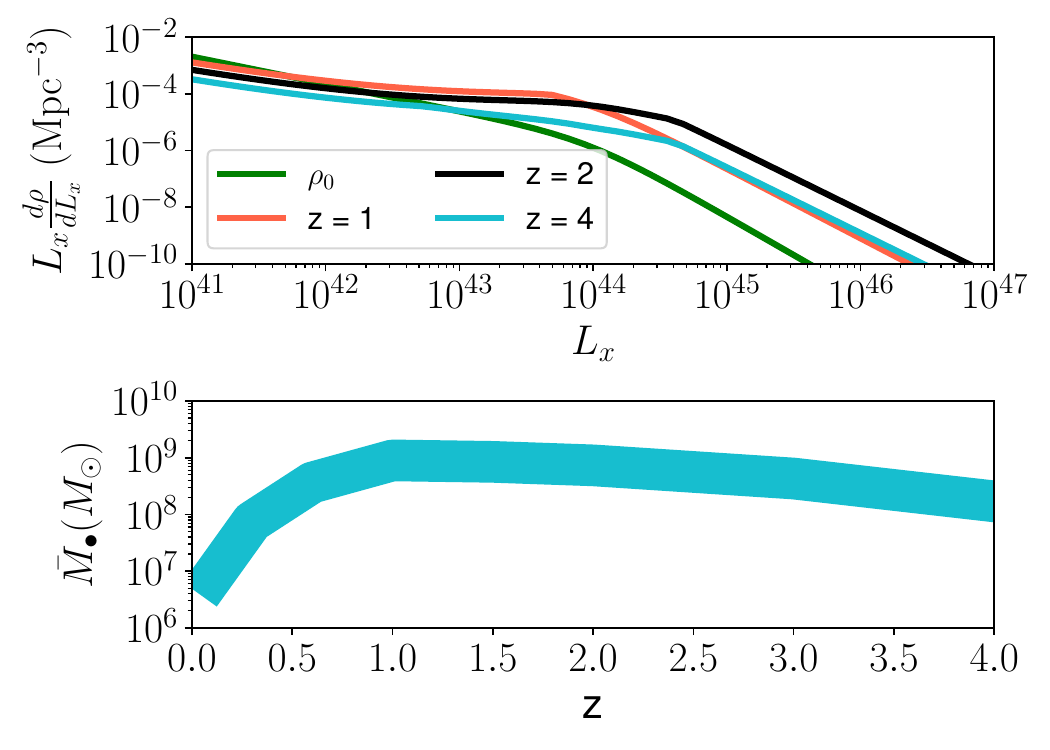}
	\caption{Upper Panel: X-ray luminosity function of active galactic nuclei used in this work \citep[See Eqn~14-19 in][]{ueda+14}.
	Lower Panel: Redshift-dependent distribution of most likely masses of active SMBH extracted from the BASS survey and the mass functions of distant actively accreting SMBH.}
	\label{fig:rho}
\end{figure}

The most defining feature of AGNs appears to be X-ray emission emanating from coronal regions \citep{padovani+17}.
However, the X-ray luminosity function cannot constrain a corresponding SMBH population because of the impact of the Eddington ratio $\lambda_E$, such that $\log{\lambda_E} = \log{\frac{L_{\rm bol}}{\rm \ergs}} - \log{\frac{M_\bullet}{M_{\odot}}} - 38.18$ \citep{ananna+22}, where $\lx (2-10 \rm keV) \simeq \lbol/20$ \citep{marconi+04}.
We hence exploit mass functions of SMBHs of AGNs to extract the most likely active SMBH mass at each redshift bin. For redshifts $z<0.3$, we use results from BAT AGN Spectroscopic Survey (BASS), and for higher redshifts, we rely on distant actively accreting SMBHs, that reside in quasars observed with the Large Bright Quasar Survey, the Bright Quasar Survey, and the Fall Equatorial Stripe of the Sloan Digital Sky Survey \citep{vestergaard+09}. The obtained mass dependence is shown in the lower panel of Fig.~\ref{fig:rho}. We then compare these results with data from \citet{merloni+08}, where we extract the most likely active black hole masses in a similar fashion. We find that both surveys yield similar black hole masses. 
Upon comparing the impact of results from \citet{vestergaard+09} and \citet{merloni+08}, we find little difference in the final neutrino spectra, especially for larger $\lx$ at higher redshifts, $\kappa_{p\gamma} \simeq 1$ (See Eq.~\eqref{eq:Lnu_Lx}). We plot an averaged form of these studies in the lower panel of Fig.~\ref{fig:rho} over an extended redshift range.

\balance

\end{document}